\documentclass[%
reprint,
superscriptaddress,
 amsmath,amssymb,
 aps,
 prl,
]{revtex4-1}

\usepackage{color,soul}
\usepackage{graphicx}
\usepackage{dcolumn}
\usepackage{bm}
\usepackage{hyperref}
\usepackage{comment}
\usepackage[mathlines]{lineno}

\begin{document}


\title{Silicon liquid structure and crystal nucleation \\from \textit{ab-initio} deep Metadynamics}

\author{Luigi Bonati}%
\affiliation{Department of Physics, ETH Zurich, c/o Universit{\`a} della Svizzera italiana, Via Giuseppe Buffi 13, CH-6900, Lugano, Switzerland}
\affiliation{Facolt{\`a} di Informatica, Instituto di Scienze Computazionali, National Center for Computational Design and Discovery of Novel Materials (MARVEL), Universit{\`a} della Svizzera italiana (USI), Via Giuseppe Buffi 13, CH-6900, Lugano, Switzerland}

\author{Michele Parrinello}%
\email{parrinello@phys.chem.ethz.ch}
\affiliation{Department of Chemistry and Applied Biosciences, ETH Zurich, c/o Universit{\`a} della Svizzera italiana, Via Giuseppe Buffi 13, CH-6900, Lugano, Switzerland}
\affiliation{Facolt{\`a} di Informatica, Instituto di Scienze Computazionali, National Center for Computational Design and Discovery of Novel Materials (MARVEL), Universit{\`a} della Svizzera italiana (USI), Via Giuseppe Buffi 13, CH-6900, Lugano, Switzerland}


\begin{abstract}
Studying the crystallization process of silicon is a challenging task since empirical potentials are not able to reproduce well the properties of both semiconducting solid and metallic liquid. On the other hand, nucleation is a rare event that occurs in much longer timescales than those achievable by ab-initio molecular dynamics.
To address this problem, we train a deep neural network potential based on a set of data generated by Metadynamics simulations using a classical potential. We show how this is an effective way to collect all the relevant data for the process of interest.
In order to drive efficiently the crystallization process, we introduce a new collective variable based on the Debye structure factor. We are able to encode the long-range order information in a local variable which is better suited to describe the nucleation dynamics.
The reference energies are then calculated using the SCAN exchange-correlation functional, which is able to get a better description of the bonding complexity of the Si phase diagram.
Finally, we recover the free energy surface with a DFT accuracy, and we compute the thermodynamics properties near the melting point, obtaining a good agreement with experimental data. In addition, we study the early stages of the crystallization process, unveiling features of the nucleation mechanism.

\begin{description}
\item[PACS numbers]
05.10.-a, 31.15.xv, 64.70.D-, 34.20.Cf
\end{description}

\end{abstract}

\maketitle


Silicon is one of the most important elements, both from the scientific and technological point of view. It is a tetravalent bonded semiconductor in the solid state, while it forms a metallic liquid with a loosely packed arrangement resulting from the persistence of covalent bonding \cite{Stich1991}.
This competition between covalent and metallic behaviour has proved to be rather difficult to model with effective potentials, as they have difficulty in describing both phases with similar precision. At the very beginning of ab-initio molecular dynamics (AIMD), the development of the Car-Parrinello method allowed providing insight into the melting mechanism, showing how the liquid is characterized by bond-breaking/bond-forming process that gives rise to its unusual behavior \cite{Stich1991}.
Since then, many studies have been conducted on the properties of both solid and liquid Si, including also \textit{ab-initio} estimates of the melting temperature \cite{Sugino1995,Dorner2018}, but a description of the crystallization process from first-principles still represents a major challenge.

Understanding the mechanism behind crystal nucleation is a long-standing goal of physics and of material science.
Classical nucleation theory suggests that thermal fluctuations in the supercooled liquid lead to the formation of a crystalline nuclei. Once the size of these nuclei reaches a critical threshold a macroscopic crystal phase is rapidly formed \cite{Sosso,Kelton2010}. Since this process takes places at the atomic or molecular scale, it is very challenging to probe experimentally. Computational studies, and particularly molecular dynamics (MD) simulations, can play an important role in revealing the atomistic details of crystal nucleation.
Unfortunately, one is hampered by the timescale problem. The free energy barrier for creating a liquid-solid interface makes the nucleation a rare event, which takes place on macroscopic time scales which cannot be reached even by the most powerful computers \cite{Valsson2016}.

A number of simulation studies based on classical potentials have been conducted. In these studies the time scale barrier has been tackled either by using deep quenches \cite{Beaucage2005,Nakhmanson2002a} that accelerate the crystallization process but might also alter its dynamics \cite{Trudu2006}, or by using enhanced sampling methods \cite{Li2009}. However, even with the use of enhanced sampling the computational cost is still too high for a fully AIMD approach. On the other hand the bonding complexity of the systems strongly suggests the use of an \textit{ab-initio} description.

An alternative that provides a good compromise between the accuracy of DFT and the efficiency of empirical potentials is offered by machine learning (ML) techniques \cite{Behler2016}. In particular Neural Networks (NN) \cite{Behler2007b} and Gaussian Process Regression \cite{Bartok2010b} have been applied to the task of creating force-fields for condensed-matter systems. The ability of ML methods to fit complex high dimensional functions has been exploited to represent the potential energy surface  as a function of the atomic coordinates $E=E(\{R_1,R_2,...,R_N\})$. One first generates a large set of configurations, computes the relative energy and forces with an \textit{ab-initio} method, and then optimizes the ML algorithm in order to get an accurate representation of the reference quantum mechanical data.

Machine-learning force fields have been already used for studying Silicon, starting from the work of Behler and Parrinello on bulk Si \cite{Behler2007b,Behler2008}. Recently also the atomistic structure of the amorphous phase has been investigated with similar methods. \cite{Deringer2018c}. In a more ambitious project a general purpose potential, based on PW91 DFT data, has been developed \cite{Bartok2018}. Our objective is less ambitious as we focus on a specific physical process, namely crystallization. What we loose in generality we hope to gain in targeted accuracy.

In order to describe the electronic structure of Si we choose to use SCAN \cite{Sun2015}, a novel exchange and correlation (XC) functional. The advantage of this functional is that being of the meta-GGA type it has a relatively modest computational cost. A feature that has attracted our attention is the ability of SCAN describe well the difference in energy between the covalent and the metallic high pressure $\beta$-tin structure \cite{Sun2016}. The latter can be seen as an idealized model for the kind of bonding that is expected in the liquid phase. Furthermore the $\beta$-tin to diamond energy difference has been argued to be correlated with the melting temperature \cite{Dorner2018}. 
The estimate of this energy difference by SCAN is improved with respect to other DFT local and semi-local XC functionals \cite{Sun2016}.

In order to build the force-field, we use the Deep Potential Molecular Dynamics (DeePMD) scheme developed by Zhang et al. \cite{Zhang2018d,Han2018}, which has a design similar to the first one proposed by Behler and Parrinello. Both schemes use neural networks (NNs) to represent the PES, which is written as the sum of atomic energies, determined by the local environment. DeePMD builds for every atom a local coordinate frame in order to preserve all the natural symmetries. We refer the readers to Ref. \cite{Zhang2018d} for further details.

Once the architecture of the neural network has been defined, we have to optimize the parameters of the NN based on a reference dataset, the so-called training set. The choice of these configurations is a crucial step. Even if NNs are able to handle the complexity of quantum-mechanical data, as they are good interpolators, they cannot predict the energies of structures  which are distant from the ones used for training \cite{Behler2016}. 
Usually, this set is composed of configurations from AIMD simulations, together with zero-temperature structures with randomly displaced atoms, and configurations whose energetic might be relevant for the process of interest \cite{Behler2015a}, starting from all the different phases involved \cite{Khaliullin2011}.
In the nucleation process interfacial energies play an important role, and it would be unwise to estimate them from the solid and liquid configurations only, where no such information is present.
This task might be even harder if one does not know the final structure the system is going to crystallize into, or whether different phases are involved in the process. In this case, one should include every different crystal structure and the relative interfaces and defects in the training set.

In order to address this problem, we propose to identify the relevant configurations from classical simulations, with the help of enhanced sampling techniques. Then we compute their energy and forces with DFT calculations using SCAN, and finally we use this reference dataset to train the DeePMD potential. Thanks to the intrinsic scalability given by the energy decomposition \cite{Zhang2018d}, once the NN has been trained on a relatively small system we can use it to study bigger system sizes and out-of-equilibrium processes like crystallization that could not be investigated otherwise with DFT accuracy.

In order to collect the reference dataset, we use Metadynamics (MetaD) \cite{Laio2002} in its Well-Tempered variant (WTMetaD) \cite{Barducci2008}.
This is an enhanced sampling method that is based on the identification of appropriate order parameters or collective variables (CVs) that describe the slow modes of the process. A repulsive potential that is function of the chosen CVs is built on the fly in an iterative process that has rigorously been shown to converge \cite{Dama2014} to the free energy surface (FES) expressed as a function of the CVs. In this way the system is pushed out of metastable states and large energy barriers can be overcome.

In the present context, MetaD can be viewed also as an efficient method to select the relevant configurations, which occupy a small portion of the configurational space. These are the ones located at the free energy minima and the states between them.
The use of WTMetaD acts as a filter that avoids including in the training set configurations that are not relevant to the phenomenon under study. This procedure is in line with the philosophy underpinning our work. Namely, we want to build a potential that is apt at describing the nucleation process rather than an all purpose potential.
So far, the use of MetaD as a tool to collect the training structures has been applied only to very small systems. Unfortunately, the procedure put forward in \cite{Herr2018} cannot be extended to condensed matter systems.

In the case of crystallization several CVs have been suggested \cite{Giberti2015}.
We report here only two paradigmatic examples. The classical work of Frenkel \cite{VanDuijneveldt1992} proposed to use the Steinhardt order parameters, which are a description of local bond order in terms of spherical harmonics \cite{Steinhardt1983}. Then the CV is built as the average of these local quantities. More recently Niu \textit{et al.} have taken a different point of view and employed the intensity of the main peak of the Debye scattering function, which is a way to enforce the coherence inherent to crystalline order \cite{Niu2018}.

\begin{figure}[b]
\includegraphics[width=\columnwidth]{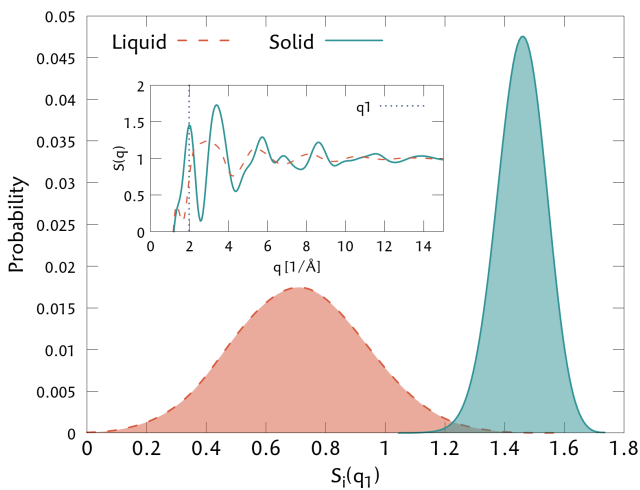}
\caption{Distribution of the local structure factor $S_i(q_1)$ in the liquid and the solid phase. Inset: Debye Structure factor computed with a cutoff radius of 8 \AA, the dashed line marks the peak located at $q_1$ used for the calculation.}
\label{fig:local-sk}
\end{figure}

Here we use a variant of Niu \textit{et al.} CV in order to further improve its ability to describe nucleation and to identify the atoms that have undergone the transition.
This can be achieved by separating the Debye equation for the structure factor \cite{Debye1915} into individual atomic contributions:

\begin{equation}
  S(q)=1+\frac{1}{N}\sum_{i=1} ^{N}\sum_{j\neq i} ^{N} \frac{\sin(q\,r_{ij})}{q\,r_{ij}}=\frac{1}{N}\sum_i ^N S_i(q)
  \label{eq:sk}
\end{equation}

In this way we can assign to every atom its own structure factor $S_i$ defined as:

\begin{equation}
S_i(q)=1+\sum_{j\neq i} ^{N_{n}} \frac{\sin(q\,r_{ij})}{q\,r_{ij}}
\label{eq:local-sk}
\end{equation}

where the sum is over all the $N_{n}$ neighbors of atom $i$ which are contained in a sphere of radius $r_c$. As in Ref. \cite{Gutierrez2002,Lin2006a} we add a damping function to alleviate the termination effects due to the finite cutoff. Details are reported in the Supplemental Material (SM).

We find out that the value of $S_i(q_1)$, where $q_1$ is the solid phase first peak, is able to distinguish clearly between solid-like and liquid-like particles.
This suggests the use of the local structure factor as fingerprint (see also Fig. SM-4 for an example) and the use of a CV which counts the number of crystal-like atoms. From Fig. \ref{fig:local-sk} we identify the atoms with $S_i(q_1)\ge \bar{q}$ as solid-like. In our case we choose $\bar{q}=1.25$.
Employing the local structure factor instead of the global one inherits the ability of $S(q)$ of discriminating the structures, but it describes also the locality typical of the nucleation.

Using the number of solid-like particles as CV (see SM) we performed a Well-Tempered MetaD simulation that is driven by the Stillinger-Weber (SW) classical potential, which is not perfect but gives a reasonably balanced description of the solid and the liquid phase \cite{Stillinger1985}. We use a system of 216 atoms and run isothermal-isobaric simulations \cite{Parrinello1981,Bussi2007,Martyna1994a} at ambient pressure and experimental melting point. The SW potential by construction reproduces well this temperature. MD simulations have been performed using LAMMPS \cite{plimpton2007lammps}, coupled with PLUMED2 \cite{Tribello2014} for the calculation of the CVs and for the MetaD bias. Additional details can be found in the SM.

\begin{figure}[t]
\includegraphics[width=\columnwidth]{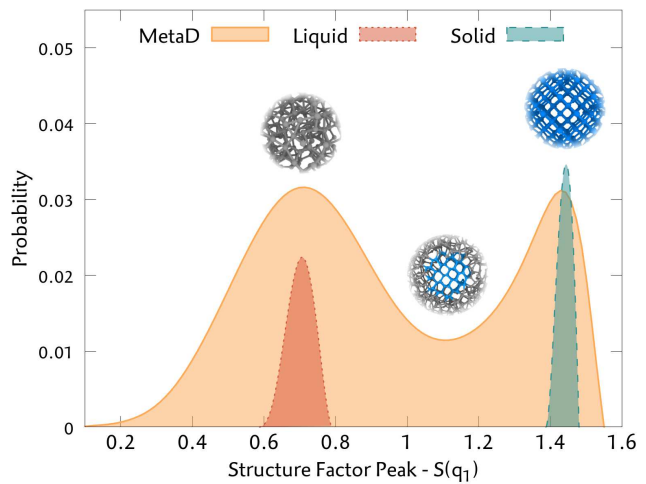}
\caption{Distribution of the configurations as a function of the structure factor intensity of the first peak, in a Well-Tempered Metadynamics simulation at 1700K. The distributions from MD simulations in the liquid and the solid are also reported. All the distributions are normalized by the number of points, solid and liquid are also scaled by a factor of 10. Examples of the structures found in the simulation are reported above the distributions.}
\label{fig:metad_histo}
\end{figure}

In Fig. \ref{fig:metad_histo} we illustrate how WTMetaD is able to explore a large number of configuration while standard MD sample a limited number of conformations of the metastable solid and liquid states only.
Of particular value for the study of nucleation is the large number of conformations harvested in the transition region where crystalline nuclei, solid-liquid interfaces and defective structures appears. This wealth of disparate but relevant configuration provides an ideal training set for the purpose of studying crystallization.

From the Metadynamics simulations training configurations are extracted every 50 fs. This ensures that they are statistically uncorrelated. By construction these configurations will be distributed as in Fig. \ref{fig:metad_histo}. We then compute the corresponding energy and forces from first-principles simulations, using SCAN. Electronic structure calculations are performed using the CP2K software with the setup reported in SM \cite{Kohn1965,Vandevondele2005,Hutter2014,Goedecker1996,Hartwigsen1998,Krack2005,VandeVondele2007,Marques2012}.

We converge the calculations with respect to k-points sampling and to energy cutoff for both the cubic diamond and the b-Sn structures, obtaining an accuracy lower than 1 meV/atom. We find that it is particularly important to use a dense k-points grid to reproduce correctly the metallic properties, despite the relatively large system size.

Then, the potential is trained using the DeePMD-kit package \cite{Wang2018} (the architecture and the details of the optimization are reported in SM). The root mean square errors (RMSE) on the testing set are equal to 2.1 meV/atom for the energies and 130 meV/A for the forces.
It is remarkable that the error made in the intermediate configurations is only slightly larger than in the equilibrium solid and liquid configurations (Fig. SM-2).
The error in the energy is of the same order of magnitude of the DFT accuracy in spite of the fact that the energy range covered (almost 1 eV/atom) is very wide. This imply that both phases are described with a similar accuracy. Also the agreement between the radial distribution functions obtained with DeePMD and AIMD references is remarkable (Fig. SM-3).

The results are robust with respect to the architecture of the neural networks and converge quickly with respect to the number of training configurations (Fig. SM-1). Furthermore, as discussed in the SM, the generality of the training set can be assessed using an ensemble of potentials, and improved with new configurations if needed.

\begin{figure}[t]
\includegraphics[width=\columnwidth]{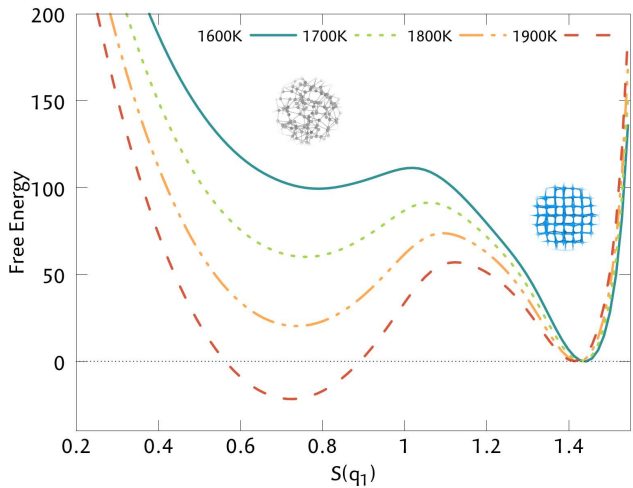}
\caption{Gibbs free energy as a function of the intensity of the first peak of the structure factor, for different temperatures around the melting point. The energy scale is in units of $k_B T_m$. The free energy profiles have been determined using the reweighting procedure of ref. \cite{Tiwary2015}. They are computed from simulations of a 216 atoms system. Finite size effects are not included, and thus the values of the barriers have to be thought as lower bound estimates.}
\label{fig:fes}
\end{figure}

Once we have trained our Metad-based NN potential we can study the crystallization process. From the many solidification and melting processes observed in the simulations, we can reconstruct an \textit{ab-initio} free energy surface at different temperatures around the melting point, showing how the relative stability between the liquid and the solid changes with the temperature (figure \ref{fig:fes}).

\begin{table}[b]
  \def\arraystretch{1.5}
  \setlength{\tabcolsep}{0.8em}
\begin{tabular}{cccc}
\hline
                          & SW   & DeePMD & EXP  \\ \hline
$T_M$ {[}$K${]}              & 1705 & 1855 & 1685 \\
$\Delta S_{sl}$ {[}$k_B${]}      & 2.39 & 3.69 & 3.59 \\
$\Delta H_{sl}$ {[}$k_B T_M${]} & 2.22 & 3.71  &  3.58  \\ \hline
\end{tabular}
\caption{Thermodynamic properties at the melting point, for the SW potential and the Metad-based NN potential trained on SCAN DFT data. Experimental data are taken from \cite{Chase1982}.}
\label{table:properties}
\end{table}

In addition, this allows computing the entropy and enthalpy difference upon phase transition (see table \ref{table:properties}). The agreement of these quantities with the experimental data is much better with respect to the one obtained with Stillinger-Weber, even if the latter was explicitly parametrized to reproduce the melting temperature and the radial distribution function of the liquid. The melting temperature that we find for the DeePMD scheme is very close to the one reported for the SCAN XC potential \cite{Dorner2018}.
This value overestimates the experimental one by 10\%, indicating that practical XC functionals still need to be improved.

\begin{figure}[t]
\includegraphics[width=\columnwidth]{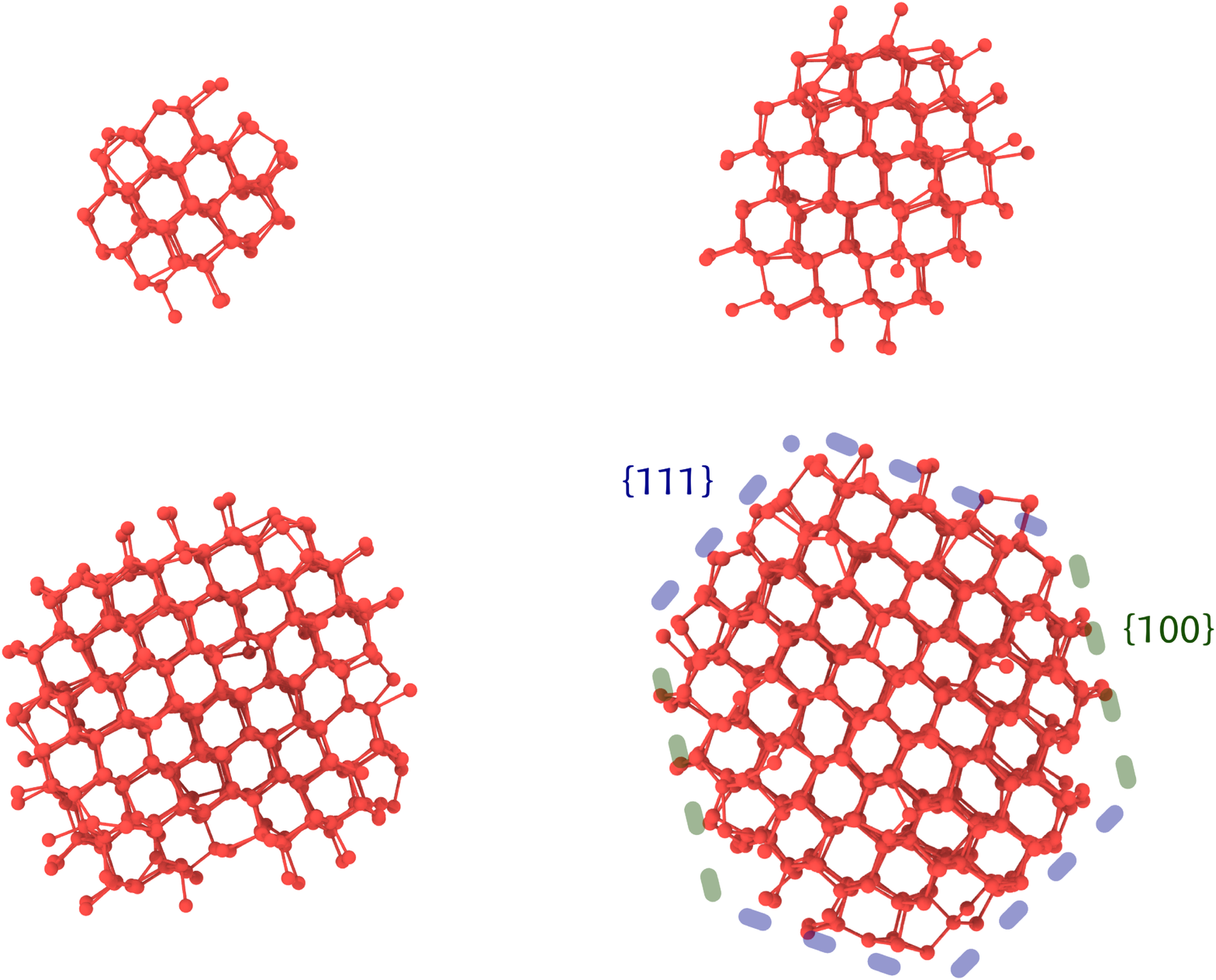}
\caption{Snapshots of the biggest cluster growing from a MetaD simulation at 1700K using the DeePMD potential. The number of atoms in the cluster is (left to right, bottom to top): 100, 200, 550 and 680. The bottom right one is close to the critical nucleus at this temperature. These have been identified using a clustering technique, where the crystallinity order parameter is the local structure factor. Then they have been analyzed with the visualization software OVITO \cite{Stukowski2010}. }
\label{fig:clusters}
\end{figure}

It is possible to investigate also the dynamical properties of liquid Si. As an example we report the value of the self-diffusion coefficient, computed from the asymptotic limit of the atomic mean square displacement. We find a value of $D=2.35 \cdot 10^{-4}\, \text{cm}^2 \text{s}^{-1}$ in good agreement with the indirected experimental measure reported in Ref. \cite{Sanders1999} of $(4.0\pm 0.5)\cdot 10^{-4}\, \text{cm}^2 \text{s}^{-1}$. The measurements are related to the melting temperature of the DPMD force field and the experimental one. For comparison we report also the value obtained with the SW potential at its melting point, which is $0.69\cdot 10^{-4}\, \text{cm}^2 \text{s}^{-1}$ \cite{Broughton}.

We are also able to follow the early stages of the crystal nucleation, observing  the formation of the clusters that eventually lead to crystallization. To do so we need to simulate large systems, in order to avoid the nuclei interacting with their periodic images. Then we use a clustering technique to identify the clusters in the system \cite{Tribello2017}, using the local structure factor as a fingerprint. In figure \ref{fig:clusters} we reported a few snapshots of such a process.
The shape of the clusters is rounded, and the eigenvalues of the inertia tensor are very close to one another.
However especially in the larger cluster crystalline facets can be observed and indexed as in the bottom right example of fig. \ref{fig:clusters}.

As a comment to the results, we believe that an \textit{ab-initio} based approach can better describe the complex conformations that are present in nucleation, with respect to empirical force-fields like SW. Moreover with the improvement in exchange-correlations functionals our approach can be made systematically more accurate.
Not to mention the fact that potentials for multi-component systems can be easily constructed. In addition, in this way one has an understanding of the DFT error and a control over the trustworthiness of the results.

In this paper we have shown how Metadynamics can be used as an effective tool that selects the relevant configurations to train neural network-based potentials for studying rare events.
This approach can be applied to condensed matter systems, and to reactive events and biophysical systems as well, where the training set might be even harder to design with a more standard approach. In addition, we have shown how the long-range order information given by the structure factor can be encoded into a local variable that can be used to drive efficiently the nucleation process. This represents a promising avenue for studying crystal nucleation in solution, where a global parameter cannot be used to drive the process.

\section*{Acknowledgements}

The authors thanks Prof. Roberto Car, Linfeng Zhang, Pablo M. Piaggi, Michele Invernizzi, Haiyang Niu, Dan Mendels, GiovanniMaria Piccini and Daniela Polino for useful discussions.
L.B. thanks in particular Pablo M. Piaggi for his guidance and encouragement, Linfeng Zhang for the precious help in using the DeePMD scheme and Michele Invernizzi for providing his implementation of the Structure Factor CV.
The research was supported by the European Union Grant No.
ERC-2014-AdG-670227/VARMET. We also acknowledge
funding from NCCR MARVEL, funded by the Swiss National
Science Foundation. Calculations were carried out on the
Mönch cluster at the Swiss National Supercomputing Center
(CSCS) and on the Euler cluster of ETHZ.

\bibliographystyle{apsrev4-1}
\bibliography{bibliography}

\end{document}